\documentclass[manuscript]{emulateapj}
\usepackage{graphicx}
\usepackage{xfrac}
\usepackage{subfigure}
\usepackage{color}
\usepackage{float}
\usepackage{amsmath}

\shorttitle{Line Ratios for Solar Wind Charge Exchange with Comets}
\shortauthors{Mullen et al.}

\begin{document}

\title{Line Ratios for Solar Wind Charge Exchange with Comets}
\author{P. D. Mullen$^{1,2}$, R. S. Cumbee$^{1,3}$, D. Lyons$^1$, L. Gu$^4$, J. Kaastra$^4$, R. L. Shelton$^1$, and P. C. Stancil$^1$}
\affil{$^1$Department of Physics and Astronomy and the Center for Simulational Physics, University of Georgia, Athens, GA 30602 \\
$^2${Department of Astronomy, University of Illinois at Urbana-Champaign, Urbana, IL, 61801}\\
$^3${NASA Goddard Space Flight Center, Greenbelt, MD, 20771}\\
$^4${SRON Netherlands Institute for Space Research, Sorbonnelaan 2, 3584 CA Utrecht, The Netherlands}}

\begin{abstract}
Charge exchange (CX) has emerged in X-ray emission modeling as a significant process that must be considered in many astrophysical environments -- particularly comets.  Comets host an interaction between solar wind ions and cometary neutrals to promote solar wind charge exchange (SWCX).  X-ray observatories provide astronomers and astrophysicists with data for many X-ray emitting comets that are impossible to accurately model without reliable charge exchange data.   Here, we utilize a streamlined set of computer programs incorporating multi-channel Landau-Zener theory and a cascade model for X-ray emission to generate cross sections and X-ray line ratios for a variety of bare and non-bare ion single electron capture (SEC) collisions.  Namely, we consider collisions between the solar wind constituent bare and H-like ions of C, N, O, Ne, Na, Mg, Al, and Si and the cometary neutrals H$_2$O, CO, CO$_2$, OH, and O.  To exemplify the application of this data, we model the X-ray emission of Comet C/2000 WM1 (linear) using the CX package in SPEX \citep{gu} and find excellent agreement with observations made with the \textit{XMM-Newton} RGS detector.  Our analyses show that the X-ray intensity is dominated by SWCX with H while H$_2$O plays a secondary role.  This is the first time, to our knowledge, that CX cross sections have been implemented into a X-ray spectral fitting package to determine the H to H$_2$O ratio in cometary atmospheres.  The CX data sets are incorporated into the modeling packages SPEX \citep{gu} and \textit{Kronos} \citep{mullen}.
\end{abstract}

\section{Introduction}
With the current orbiting X-ray observatories and plans for future high-resolution X-ray missions, the need for atomic and molecular data is universally recognized by astrophysical modelers. For instance, with the possible \textit{Hitomi} detection of charge exchange (CX) in the Perseus cluster or the recent \textit{Chandra} observations of X-ray emission from Pluto, accurate atomic data are necessary to understand such unexpected, very recent observations \citep{gupers, pluto}.  

Another interest in CX X-ray modeling is the study of comets (such as Comet C/2000 WM1 linear, Comet Tempel 1, etc.) whose primary mechanism for X-ray emission is solar wind charge exchange (SWCX). Cometary SWCX occurs when highly charged projectile ions, such as bare and H-like C, N, O, Ne, Na, Mg, Al, and Si, present in the solar wind, capture an electron from a target neutral species, such as H$_2$O, CO, CO$_2$, OH, O, and H, present in the cometary atmosphere.  Although more than one electron can be captured  during a charge exchange collision, we only consider single electron capture (SEC) in this work.  The SEC process is given in Equation (1) where $X^{q+}$ denotes the projectile ion and $Y$ denotes the collision target, 
\begin{equation}
X^{q+} + Y \rightarrow X^{(q-1)+} (n \ell \; ^{2S+1}L) + Y^{+} + \Delta E,
\end{equation}
while $n \ell$ specifies the principal quantum number and orbital angular momentum of the captured electron, $^{2S+1}L$ denotes the total spin $S$ and total orbital angular momenta $L$ of the product ion, and $\Delta E$ gives the kinetic energy liberated from the collision system.  During cometary SWCX, electrons are captured into highly energetic states of the solar wind ions.  To stabilize, excess internal energy is emitted in the form of X-ray photons so that the captured electron is demoted to the ground state. This charge exchange induced X-ray emission is observed by X-ray observatories such as \textit{Chandra}, \textit{XMM-Newton}, and \textit{Suzaku} \citep[e.g.,][]{lis96,cra00,6,bha07,suda, 10}, but can also be detected in the laboratory, through experimental apparatuses such as an electron beam ion trap \citep{1, war2}.  In this work, we attempt to model such observations by applying 1) multi-channel Landau-Zener (MCLZ) theory to generate $n \ell S$-resolved cross sections for charge exchange and 2) a cascade model to generate theoretical X-ray spectra and ultimately, line ratios.  These line ratios can be implemented into X-ray spectral fitting packages to model cometary environments and be used as a diagnostic for relative abundances of neutrals in cometary atmospheres while also putting constraints on solar wind composition and velocities at the comet.

We compare our theoretical atomic and molecular data to other available data in the literature such as other cross sections obtained by different theoretical means, experimental X-ray spectra, or even available line ratios themselves.  Such comparisons appear in the Appendix.   These comparisons are limited, however, as to our knowledge, this is the first time that such calculations have been performed for many of the aforementioned systems.  

\section{Theory}
The MCLZ theory applied to generate $n \ell S$-resolved cross sections in this work is detailed in \cite{mullen} and \cite{lyo17}.  To briefly summarize the schema, we begin by considering all possible initial and final channels of the collision system by applying Wigner-Witmer rules \citep[i.e. angular momentum/molecular symmetry conservation,][]{herz,wigner}.  
\begin{figure}[H]
\centering
\includegraphics[width=0.48 \textwidth]{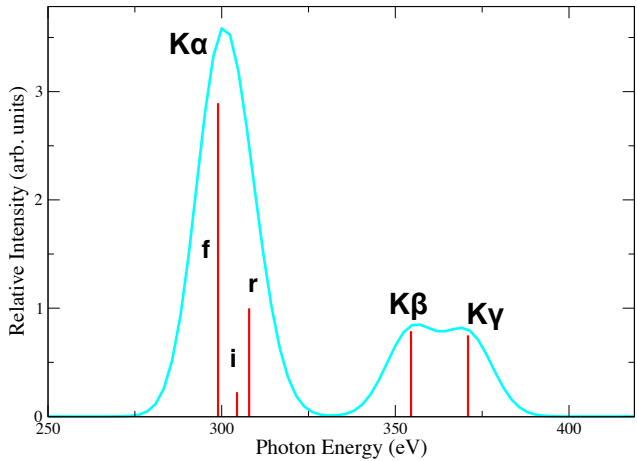}
\caption{Theoretical MCLZ/cascade model spectrum for C$^{5+}$ collisions with H$_2$O for a collision energy of 1 keV/u.  The spectra is normalized to the K$\alpha$ resonance (r) line. The K$\alpha$ forbidden and intercombination
lines are labeled as f and i, respectively. An FWHM resolution of 10 eV was convolved with the theoretical line spectrum to create the blue curve.}
\label{fig1}
\end{figure}
\noindent The NIST Atomic Spectra Database \citep{9} then provides product ion excitation energies necessary for constructing the potentials associated with allowed final channels.  Missing states in the NIST Atomic Spectra Database are approximated with quantum defect theory \citep[see][]{mullen}.  The couplings between initial and final channels are characterized by the energy differences of the adiabatic potentials at internuclear distances between the projectile-collision pairs at which transitions may occur (avoided crossings).  The charge exchange probability is thus dependent upon 1) avoided crossing locations, $R_c$, 2) the energy differences at these $R_c$, and 3) the difference between the derivatives of the initial and final diabatic potential curves at $R_c$.  We obtain the charge exchange cross section by considering 1) the incident collision energy, 2) all partial waves of the collision system, and 3) the \cite{5} multi-channel probability relation.  For bare ion collisions, the {\it low-energy} $\ell$-distribution model, as shown in equation (2),
\begin{equation}
W^{\rm{Low}}_{n\ell} = (2 \ell + 1) \frac{[(n-1)!]^{2}}{(n+\ell)!(n-1-\ell)!},
\end{equation}
is applied \citep{6}.   Other $\ell$-distribution models may be applied \citep[see, for example,][]{6,7}.  Non-bare ion collisions do not require the use of an $\ell$-distribution model as the product ions of charge exchange are non-degenerate.
\linebreak
\indent
Using the resulting $n \ell S$-resolved cross sections to seed a cascade model for X-ray emission, we can track the network of emitted photons due to the evolution of cascading electrons to stability.  Thus, with the cross sections giving an initial population of states, transition probabilities (Einstein A coefficients) dictating which cascade network is most probable, and obeying transition selection rules, the cascade model yields theoretical X-ray spectra and line ratios.  As an example of the cascade model output, we present the theoretical spectrum for C V emission due to the charge exchange collision between solar wind C$^{5+}$ and the cometary neutral H$_2$O in Figure 1.  Note that all lines are normalized to the K$\alpha$ resonance (r) line. In this example, the strength of the forbidden (f) line is a characteristic of the CX X-ray
emission process as opposed to electron impact excitation where the resonance line typically dominates.

Special considerations are mandatory for bare-ion and hydrogen-like ion collisions with OH and O.  The ground electronic states for OH and O are $^2 \Pi$ and $^3P_g$, respectively.  Because of the non-zero angular momentum of these electronic states, modifications to the initial interaction potential utilized in \cite{mullen} and \cite{lyo17} are made so that the long-range interaction includes the quadrupole moment, $Q_M$, of OH and O.  The potential, therefore, is of the form, 
\begin{equation}
V_i (R) = A\exp(-BR) - \frac{\alpha q^2}{2R^4} + \frac{q Q_M}{2 R^3},
\end{equation}
where coefficients $A$ and $B$ are estimated in \cite{2} and $\alpha$ is the dipole polarizability of the neutral target and $R$ is the internuclear distance between the projectile/target pair.  Values for the quadrupole moments of O and OH are given by \cite{oquad} and the \cite{ohquad}, respectively, as $Q_{M, O} = -0.95$ a.u. and $Q_{M, OH} = -3.323$ a.u.. 

In addition to modifications to the initial channel model, Wigner-Witmer rules, based upon momentum conservation principles, require that bare ion collisions with OH conserve the $^2 \Pi$ molecular electronic symmetry; therefore, only product projectile ion states with $\ell \geq 1$ are allowed thus requiring a renormalization of the adopted $\ell$-distribution functions.  For hydrogen-like ions colliding with OH, the system must conserve the $^3 \Pi$ symmetry; therefore, only singlet and triplet states with $\ell \geq 1$ are allowed for the product projectile ion.  Resulting cross sections must be weighted by the approach probability $g$-factor (see \cite{wigner} for details) which is 1/4 for $^1 \Pi$ states and 3/4 for $^3 \Pi$ states. These considerations for OH targets assume that CX with OH results purely from linear collisions.  

For bare ion collisions with O, $^3 \Sigma^-$ and $^3 \Pi$ molecular electronic symmetries must be conserved and be weighted by the $g$-factors 1/3 and 2/3, respectively.  For this collision, doublet states of the product projectile ion conserve the $^3 \Sigma^-$ symmetry while only doublet states with $\ell \geq 1$ conserve the  $^3 \Pi$ symmetry.  Finally, hydrogen-like ion collisions with O conserve the $^2 \Sigma^-$, $^2 \Pi$, $^4 \Sigma^-$, and $^4 \Pi$  symmetries each weighted by $g$-factors 1/9, 2/9, 2/9, and 4/9, respectively.  All singlet and triplet states of the product projectile ion conserve the $^4 \Sigma^-$ electronic state while only triplet states conserve the $^2 \Sigma^-$ symmetry.  All singlet and triplet states with $\ell \geq 1$ conserve the $^4 \Pi$ symmetry while only triplet states with  $\ell \geq 1$ conserve the $^2 \Pi$ symmetry; therefore, again, the $\ell$-distribution functions must be renormalized.

\section{Line Ratios} 
As our intent is to help astronomers and astrophysicists by providing atomic and molecular data for its use in X-ray emission modeling, we present calculated line ratios for the various aforementioned cometary SWCX projectile-collision pairs at five characteristic collision velocities (200 km/s, 400 km/s, 600 km/s, 800 km/s, and 1000 km/s) in Table 1 of the Appendix.  The line ratios are defined as the ratio between an emission line and the Ly$\alpha$ line (for bare projectile ion collisions) or the K$\alpha$ resonance line (for H-like projectile collisions).  For bare projectile ion collision systems, we have applied the low-energy distribution to MCLZ $n$-resolved cross sections before inputting them into the cascade model.  Line ratios that used other distribution functions are available in the \textit{Kronos}\footnote{https://www.physast.uga.edu/ugacxdb/} database or upon request.  For H-like projectile ion collision systems, MCLZ $n \ell S$-resolved cross sections are used to generate line ratios; here, we also report forbidden, intercombination, and resonance lines for K$\alpha$ emission.   
\linebreak
\indent
The reliability of these data is tested in several ways.  First, the \cite{mullen} study of the high charge and computationally expensive/extensive Fe XXV and Fe XXVI charge exchange collision systems (with various targets) were investigated.  Subsequent analyses revealed very good agreement with two electron beam ion trap studies \citep{1, war2} as well as other theoretical calculations such as CTMC (classical trajectory Monte Carlo) as given in \cite{kats}, \cite{schultz}, and \cite{war2}.  Next, we compare MCLZ cometary SWCX calculations to available data in the literature--albeit limited.  These comparisons are detailed in the Appendix.  

\section{Application -- Comet C/2000 WM1 (linear)}
To exemplify how these results can be applied, we incorporate the charge exchange data from this work and \cite{hwork} into the charge exchange package in SPEX \citep{gu} to perform a CX fitting to Comet C/2000 WM1 (linear).  X-ray satellite \textit{XMM-Newton} \citep{xmm} observed Comet C/2000 WM1 in December 2001 (Observation ID: 0103461101).  Here, we focus on high spectroscopic resolution data taken by one \textit{XMM-Newton} instrument, the Reflection Grating Spectrometer (RGS).  The same RGS data have been previously fitted in \cite{gu}; the main difference, however, is that \cite{gu} assumes only H targets,
while in this work, many cometary targets have been tested (H$_2$O, CO, CO$_2$, OH, and O).  Only a short period (18 ks) of exposure is analyzed.  The rest of the data were collected outside the RGS field of view, which was 5 arcmin in this observation. 

In the X-ray modeling package, SPEX, we first implement the charge exchange MCLZ data for bare and H-like C to Si collisions with H and H$_2$O.  However, quantum molecular-orbital close-coupling data were adopted for bare and hydrogen-like C, N, and O collisions with H \citep{nolte, wu-stancil} as reported in \cite{10,hwork}. 

To fit the spectrum, we subtract the standard model background and correct for instrumental broadening due to the spatial extent of the diffuse comet halo and its motion.  The 18-36.5 \AA\ spectrum was fitted with two SWCX components, one with atomic H targets and the other with H$_2$O targets \citep[both naturally expected in comet spectra,][]{bode}.  The ionization temperatures of the two SW components are free to vary, while their collision velocities and metal abundances are assumed to be the same.   As shown in Figure 2, the RGS spectrum can be reasonably fit with the model.  The best-fit C statistic is 340 with 287 degrees of freedom, a significantly better fit than the value obtained in \cite{gu}.  We have tried fitting the RGS data with all molecular targets from this work (H$_2$O, CO, CO$_2$, OH, and O), but the H and H$_2$O combination gives the best fit.  The fit determines the H target temperature to be 0.13 keV.  For the water component, the current model prefers a two-temperature configuration at 0.22 keV and 0.06 keV, which might suggest the cooling of solar wind ions when penetrating into the cometary atmosphere.  The CX collision velocity is determined to be $\sim$240 km/s. We also attempt to fit the spectrum with two H-target only components.  This yielded a worse fitting by $\Delta$C = 39.  This indicates that the cometary H and H$_2$O components should both contribute to the CX emission. The emission measure of CX collisions with H$_2$O is less than 1/3 ($\sim$30\%) of that with H.  Our analyses of this fitting routine can be compared to that of \cite{bode}.  For instance, \cite{bode} predicts that the dominant target changes when the comet approaches the Sun.  As seen in  Figure 9 of \cite{bode}, Halley's Comet at 1 AU has more H in its cometary atmosphere than Hale-Bopp at 3.1 AU, due to the photodissociation of H$_2$O.  Comet C/2000 WM1 (linear) is observed at around 1 AU so it should behave more closely to that of the Halley case.  Based on the Halley investigation, the H$_2$O core of C/2000 WM1 (linear) has a diameter of $3 \times 10^4$ km (1.6 arcmin) in the observational frame, while H becomes more important in the outer coma ($1 \times 10^5 - 3 \times 10^6$ km).  Since the RGS spectrum (which has no spatial resolution) was taken of a $R \sim 4 \times 10^7$ km region centered on the comet core, we predict that the overall emission due to charge exchange with H is larger than that due to water. 

The possibility of single-capture, multiple-collision events which eventually neutralize the solar wind ion, was also tested  using a model from \citet[see][]{gu}, but this was found not improve the fit.

\begin{figure*}
\centering
\includegraphics[width=0.9\textwidth]{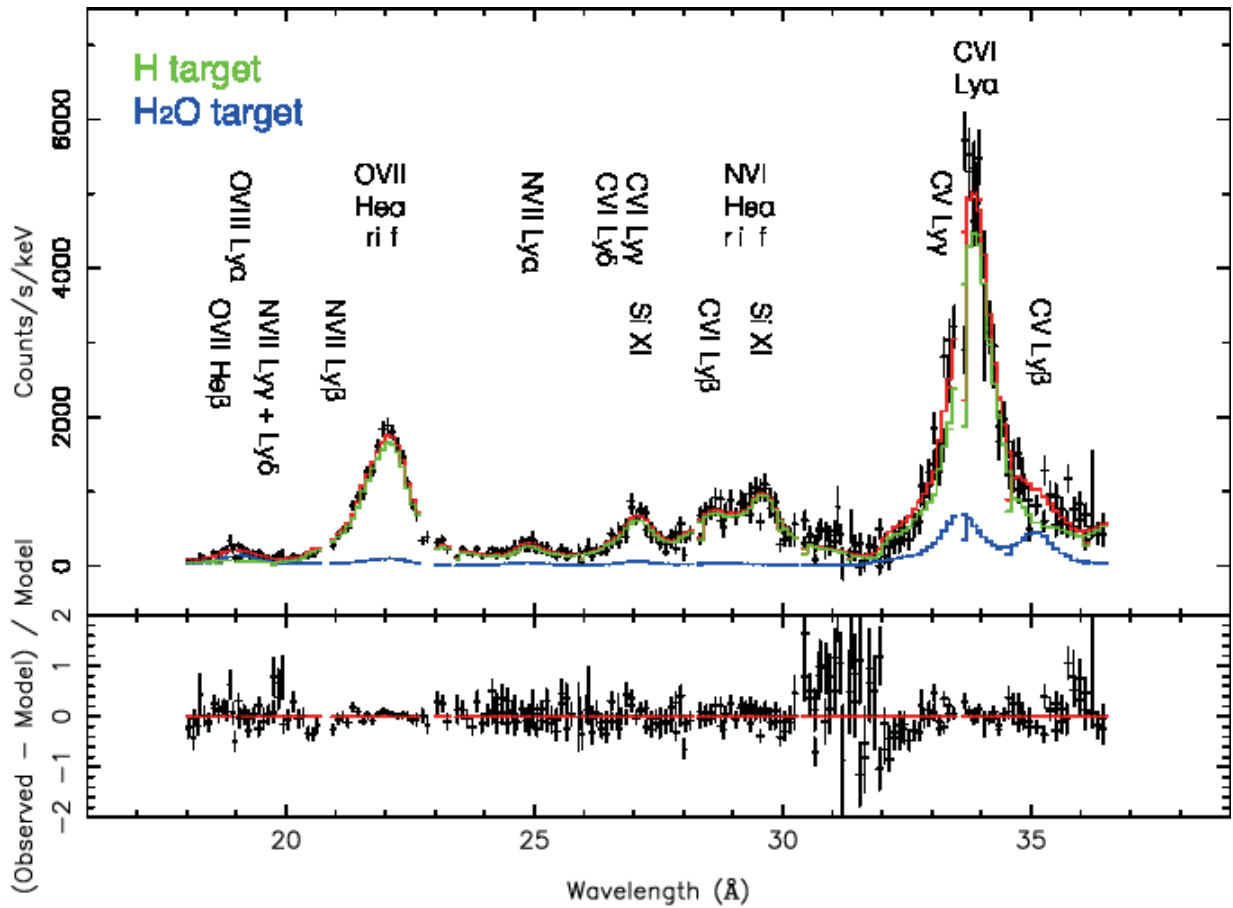}
\caption{X-ray emission modeling of \textit{XMM-Newton} observations of Comet C/2000 WM1 (linear).  The black curve and error bars correspond to observations made by the XMM-RGS detector.  The red curve gives the theoretical total X-ray emission spectrum due to CX collisions between solar wind ions (mostly C$^{5+, 6+}$, N$^{6+, 7+}$, and O$^{7+, 8+}$) and cometary gas (H and H$_2$O).  Contributions to the total spectrum due to CX collisions with H and H$_2$O are given in the green and blue curves, respectively.}
\label{fig2}
\end{figure*}

\section{Conclusions}
With current orbiting satellites, plans for future high-resolution X-ray missions underway, and the development of a streamlined MCLZ approach, charge exchange line ratios can be applied in X-ray emission models to advance the astrophysical modeling of many environments -- particularly comets.  To exemplify the need and use for such data, the modeling of Comet C/2000 WM1 linear illustrates one of the most comprehensive applications of atomic data to the astrophysical modeling of a comet to-date.  Subsequent analyses of this model show that there is excellent agreement between theory and \textit{XMM}-RGS observations and that the X-ray intensity is dominated by cometary SWCX with H as predicted in \cite{bode}.  Further, with the good agreement between our theoretical results and other available data, as shown in the Appendix, as well as the performance of the MCLZ method  for the computationally expensive Fe XXV and Fe XXVI systems discussed in \cite{mullen}, we demonstrate the reliability of this charge exchange data and its future application in high-resolution X-ray spectral modeling.  All calculated CX cross sections and line ratios appearing in this work are available 
in SPEX \citep{gu} and in the \textit{Kronos} package \citep{mullen}.  Scripts are also available upon request for implementing \textit{Kronos} line ratio data into XSPEC \citep{xspec} models.  
\\
 \acknowledgements
This work was partially supported by NASA grants NNX09AC46G and NNX13AF31G.  We thank Randall Smith for helpful discussions.

\newpage
\vspace*{-0.75 cm}
\appendix
The benchmarking of MCLZ cross sections was a major focus in \cite{mullen} due to the computational difficulties of the Fe XXV and Fe XXVI charge exchange systems.  In this work, we continue benchmarking the MCLZ code \citep[{\it Stueckelberg},][]{lyo17} by comparing to theoretical and experimental data where available.  Unfortunately, the theoretical data available for comparison are limited as, to our knowledge, this is the first time that such calculations have been performed for many of these bare and non-bare ion collisions with cometary neutral targets -- only a few cross sections exist for comparison.  Fortunately, there are some measurements of charge exchange induced X-ray emission for several of the systems in this work.  
Using the cascade model and MCLZ $n \ell$-resolved cross sections, theoretical spectra for O$^{8+}$ collisions with H$_2$O with a FWHM of $\sim$100 eV are given in Figure 3 for a collision energy of 3 keV/u.  To compare, we also present experimental spectra as obtained by \cite{greenwood} utilizing an electron cyclotron resonance source for a collision energy of 3.1 keV/u.  Note that the theoretical and experimental spectra are normalized to the Ly$\alpha$ peak.  Also note that the experimental spectrum presented in Figure 3 has not been corrected for detector efficiency.  As seen in Figure 3, the high-$n$ predicted emission using the low-energy distribution is significantly larger than that of the experiment.  This is an interesting finding as the \cite{mullen} study for Fe$^{26+}$ collisions with N$_2$ predicted a spectra that yielded lower high-$n$ emission peaks than measured by the \cite{1} EBIT study.  This is most likely due to the large difference in collision energy associated with both experiments (i.e. \cite{1}: $\sim$10 eV/u, \cite{greenwood}: $\sim$3 keV/u).  Regardless, theoretical spectra are clearly sensitive to the $\ell$-distribution function applied to MCLZ $n$-resolved cross sections.  The low-energy distribution applied in Figure 3 and in the line ratios presented in this work is expected to be applicable for collision energies less than 1 keV/u.  Because the collision energy is 3 keV/u or greater, it is possible that other distribution functions can perhaps better model the spectrum.  Therefore, we apply three other distribution functions -- namely, the {\it modified low-energy}, the {\it separable}, and the {\it statistical} distribution functions \citep[see, for example, ][]{7, mullen}.   The statistical distribution, which is expected to be applicable at very high collision energies, gives high-$n$ emission significantly less than experiment.  However, the separable distribution function, which is the only distribution function dependent on projectile ion charge, was also tested and shows decent agreement in modeling the high-$n$ emission peak.  From Figure 3, it appears that a blend of characteristics from both the separable distribution and low-energy distributions would reproduce the experimental spectrum,  but multielectron capture (MEC) effects, not included here, might also be present in the measured spectrum \citep[see, for example, ][]{ali05}.  The MEC process, double capture followed by autoionization (DCAI), tends to enhance the population of
lower $n$-states and therefore increase the relative intensity of Ly$\alpha$ and possibly Ly$\beta$.

\begin{figure}[H]
\centering
\includegraphics[width=0.6 \textwidth]{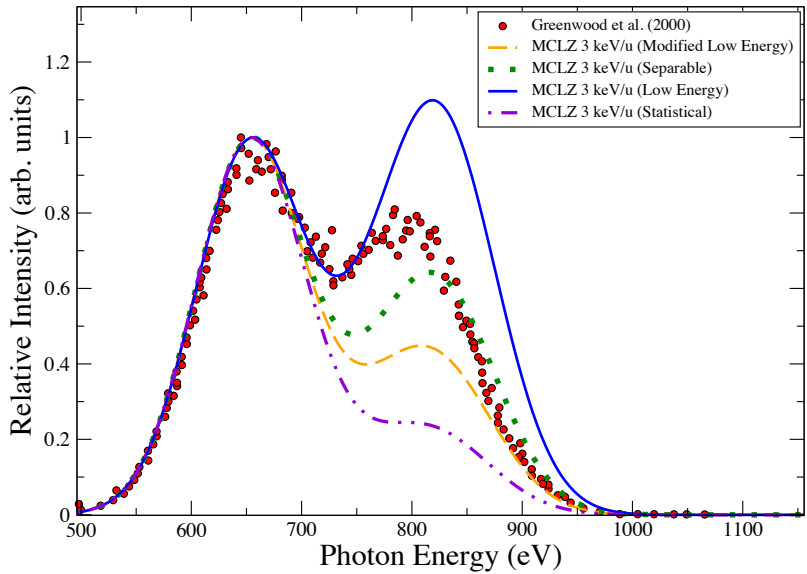}
\caption{Comparison of theoretical MCLZ/cascade model spectra to \cite{greenwood} experimental spectrum for O$^{8+}$ collisions with H$_2$O at a collision energy of 3.1 keV/u.  Each spectra is normalized to the Ly$\alpha$ peak.   MCLZ spectra utilize various $\ell$-distribution models as described in the text.  The theoretical spectra mimic a $\sim$100 eV FWHM resolution \citep{greenwood}.  Note that the experimental spectrum has not been corrected for detector efficiency.}
\label{fig3}
\end{figure}

MCLZ calculations for non-bare ion collisions require no application of a distribution function as product ion states are non-degenerate.  Therefore, comparisons of such collisions with available experimental data could give a better and more pure measure of MCLZ performance as opposed to the use of distribution functions.  Figure 4 gives the theoretical spectra resulting from Ne$^{9+}$ charge exchange collisions with H$_2$O compared to experimental spectra from the same \cite{greenwood} study.  Again, the \cite{greenwood} spectrum given in Figure 4 has not been corrected for detector efficiency.  Spectra for collision energies of 1, 2, and 3 keV/u are presented while the experimental collision energy is $\sim$2.9 keV/u.

\begin{figure}[H]
\centering
\includegraphics[width=0.6 \textwidth]{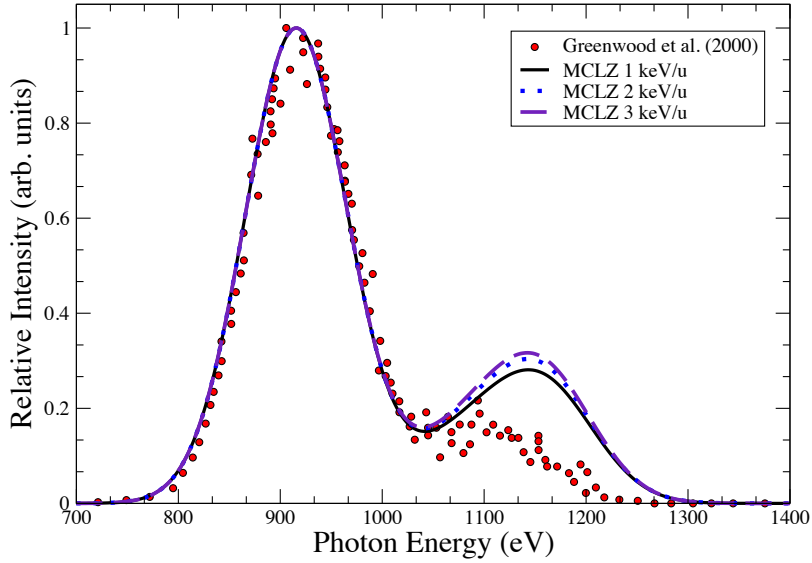}
\caption{Comparison of theoretical MCLZ/cascade model spectrum to \cite{greenwood} spectrum for Ne$^{9+}$ collisions with H$_2$O at an experimental collision energy of $\sim$2.9 keV/u.   Each spectra is normalized to the maximum intensity of the K$\alpha$ emission. The theoretical spectra mimic a $\sim$100 eV FWHM resolution.  No distribution function is applied in the MCLZ calculation as this is a non-bare ion collision.}
\label{fig4}
\end{figure}

In Figure 4, we see a slight enhancement of high-$n$ emission as compared to the \cite{greenwood} study.  This emission feature can be further investigated by looking to the theoretical line ratios -- for instance, the K$\beta$ line ratio predicted for this collision system is equal to 0.2373 while those for the K$\delta$ and K$\epsilon$ lines (5$p \rightarrow$ 1s, 6$p \rightarrow$ 1s) are 0.1206 and 0.3421, respectively, for a collision energy of 1 keV/u.   The reasoning behind these relatively high K$\delta$ and K$\epsilon$ lines can be explained by looking into the factors that dictate their intensities: 1) their associated cross section, 2) the corresponding Einstein A coefficient, and 3) the triplet-singlet cross section ratio, which is not necessarily 3:1.   

\begin{figure}[H]
\centering
\includegraphics[width=0.6 \textwidth]{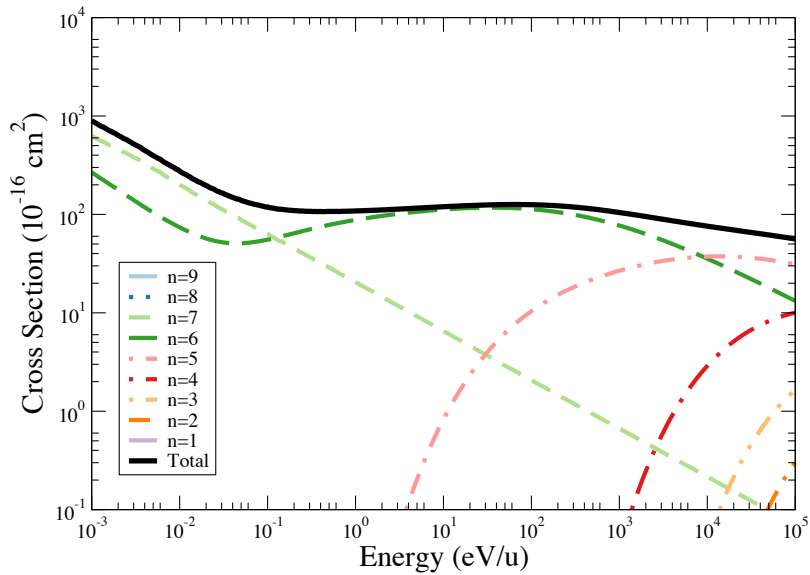}
\caption{\footnotesize SEC $n$-resolved MCLZ cross sections for Ne$^{9+}$ collisions with H$_2$O summed over $\ell$ and $S$.  Dominant $n$-channels at 1 keV/u are $n$=5, 6.}
\label{fig5}
\end{figure}

To the first, Figure 5 gives the $n$-resolved cross sections for Ne$^{9+}$ CX collisions with H$_2$O.   At the collision energies of interest, 1-3 keV/u, the dominant channels are $n$=5, 6 which is exactly as expected by the simple formula for dominant $n$-capture, $q^{0.75} \sim 5.2$, where $q$ is the charge of the projectile ion.  Accompanying these large cross sections into the $n=5, 6$ channels, the associated K$\delta$ and K$\epsilon$ Einstein A coefficients are $\sim$ 5.2$\times$10$^{11}$ s$^{-1}$ and $\sim$3.0$\times$10$^{11}$  s$^{-1}$ which are roughly of the same order of magnitude as other $np \rightarrow 1s$ transitions for $n$ less than $n=5,6$.  

Because the reported experimental spectrum in \cite{greenwood} was not corrected for the transmission efficiency of the Be window of their X-ray detector, we also give a comparison of our theoretical line ratios to \cite{greenwood} line ratios that \textit{do} correct for the detector efficiency in Tables 2 and 3.  Note that the experimental line ratios presented in Tables 2 and 3 are deduced from their reported relative contributions of transitions to the total spectrum (again, after correcting for detector efficiency).   Table 2 corresponds to line ratios for bare projectile ion CX collisions while Table 3 corresponds to line ratios for hydrogen-like projectile ion CX collisions.  \cite{greenwood} was not able to resolve independent constributions from Ly$\gamma$ and Ly$\delta$ for many of the CX collisions.  Therefore, in such instances, we report the sum of these.   In Table 2, we report line ratios utilizing four different distribution functions: \textit{low energy}, \textit{statistical}, \textit{modified low energy}, and \textit{separable}.  \cite{greenwood} were also not able to resolve the intercombination and resonance lines for He-like emission, nor do the experiments observe the forbidden line due to the long lifetime of the $2s~^3S$. Therefore, for Table 3, we normalize theoretical line ratios to the sum of K$\alpha$ f, i, and r.  Also reported in these two tables are hardness ratios.  

As seen in Table 2, the separable or statistical distribution best replicates experimental hardness ratios as expected due to the relatively high collision energies.  
An exception is the N$^{7+}$ + H$_2$O system which is best reproduced by the low-energy distribution, but this appears to be an anomaly. The low-energy distribution consistently reports larger hardness ratios for these systems though these collision energies are significantly larger ($E_{\textrm{collision}} \sim$ 3 keV/u) than those applied in the cometary model.

Table 3 demonstrates that MCLZ $n \ell S$-resolved cross sections yield larger hardness ratios than measured in \cite{greenwood}.  This is consistent with the findings of Figure 4 which demonstrated an enhancement of high-$n$ emission compared to experiment.  Again, perhaps multiple electron capture could be the source of such discrepancy.  We anticipate that the effect of multiple electron capture in experiment would enhance $n=2\rightarrow1$ emission which would therefore decrease hardness ratios and, in turn, decrease the line ratios associated with high$-n$ emission.   

Another study that we compare our theoretical data to is that of \cite{beiers} who used the same microcalorimeter spectrometer originally designed for the \textit{Astro-E} mission.   In Figure 6, we present theoretical spectra for O$^{8+}$ collisions with N$_2$ at a collision energy of 10 eV/u compared to the EBIT measurements that have an estimated collision energy of $\sim$10 eV/u.  Applying the low-energy distribution function (Equation 2), we see excellent agreement between experimental and theoretical line ratios for Ly$\alpha$, Ly$\beta$, and Ly$\gamma$ emission that the microcalorimeter has no difficulty in resolving.  However, again, for high-$n$ emission, theory predicts a much higher contribution to the spectra than the EBIT measurement.  Therefore, we look to other distribution functions -- namely the modified low-energy distribution and the separable distribution (note that this collision energy is so small that the statistical distribution is not considered).  Both the separable and modified low energy distribution model Ly$\delta$ well.  However, it is worth noting that the low energy distribution function models Ly$\alpha$, Ly$\beta$, and Ly$\gamma$ most accurately and would be expected to be valid at this low collision energy.  

\begin{figure}[H]
\centering
\includegraphics[width=0.6 \textwidth]{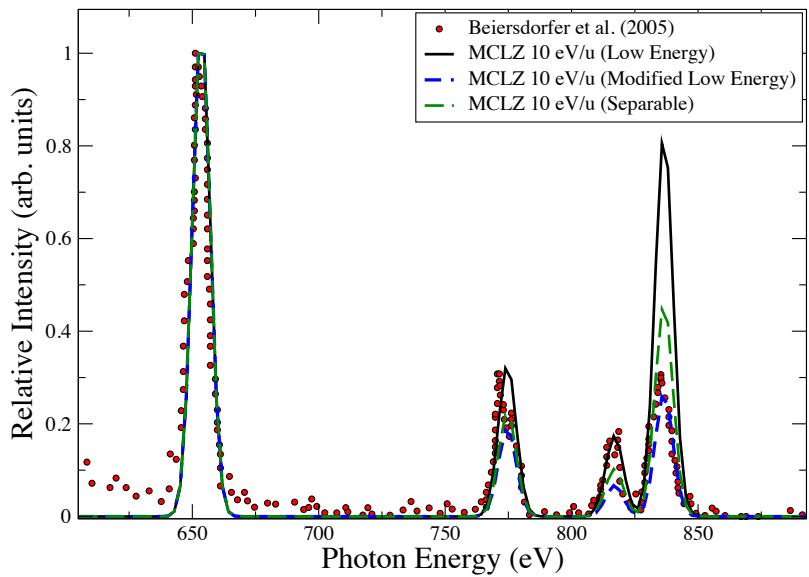}
\caption{\footnotesize Comparison of theoretical MCLZ/cascade model spectrum to the \cite{beiers} experimental spectrum for O$^{8+}$ collisions with N$_2$.  Each spectrum is normalized to the maximum intensity of Ly$\alpha$ emission. The theoretical spectra mimic a $\sim$10 eV FWHM resolution. \cite{beiers} estimate an experimental collision energy of $\sim$10 eV/u.}
\label{fig6}
\end{figure}

We benchmark MCLZ cross sections directly to other available theoretical data and extrapolations from experiment as seen in Table 4.  \cite{bigbook} shows agreement between relative $n$-resolved cross sections for various systems with previous Landau-Zener (LZ) calculations (see \cite{bigbook} for details) and CTMC calculations.  For instance, MCLZ calculations for N$^{7+}$ collisions with CO at collision energies of 2 keV/u are in almost perfect agreement with previous LZ calculations and are comparable to CTMC calculations.  Interestingly, all theoretical methods predict that $\sigma_{n=4} > \sigma_{n=5}$ whereas experiment suggests that $\sigma_{n=5}$ comprises nearly 75\% of the total cross section.  Identical behavior is seen for N$^{7+}$ collisions with CO$_2$ at collision energies of 2 keV/u -- namely, MCLZ is in agreement with CTMC and previous LZ calculations whereas experiment suggests a different dominant channel than all theoretical calculations.   This disagreement in dominant channels between theory and experiment is not universal.  Comparing MCLZ $n$-resolved cross sections for Ne$^{10+}$ + H$_2$O CX collisions at $\sim$10 eV/u to the \cite{rig02} study which extracts populations from X-ray measurements, both sets of data agree that the dominant channels are $n=6,7$.

As previously mentioned, few theoretical CX calculations for systems involving cometary molecules have been performed.  Therefore, Table 4 lacks CTMC and previous LZ calculations for all O$^{7+}$ collisions.  However, MCLZ calculations for the O$^{7+}$ + H$_2$O CX collisions at 2 keV/u, a particularly important solar wind/cometary gas interaction, are in excellent agreement with experiment.   However, we note again that the current work considers only
single electron capture, while the effects of multiple capture from multi-electron targets at SW velocities is not well understood.

Finally, we compare MCLZ line ratios for bare and H-like C, N, and O collisions with H$_2$O to \cite{bode07} in Table 5.  The \cite{bode07} data corresponds to collisions with H, \textit{not} H$_2$O.  This comparison emphasizes the target dependence of not just CX cross sections but the resulting line ratios themselves.  Thus, we suggest that X-ray modelers consider, when available, the dependencies of 1) projectile ion, 2) target, and 3) collision energy.
This is now possible due to the fact that all data presented here and in \citet{hwork} are available in
{\it Kronos} and SPEX. Further, 
scripts are available to generate line ratios in XSPEC format using recommended cross sections from {\it Kronos}. In these scripts, each X-ray emission line, i.e. Ly$\alpha$ or Ly$\beta$, is put in separately as a Gaussian parameter. Energies from 200 eV/u ($\sim$200 km/s) to 5000 eV/u ($\sim$1000 km/s) can be chosen, and ions from C to Si can be considered, using the best available cross-sections for a given ion-neutral interaction. Additional details are given in \citet{hwork}.

\newpage

\onecolumngrid
\LongTables
\begin{deluxetable*}{c c c c c c c c c c}
\centering
\tabletypesize{\tiny}
\tablecaption{Charge Exchange Induced X-ray Emission of H and He-like C-Si}
\tablewidth{0pt}
\tablehead{
\multicolumn{1}{c}{Phot. Energy} &
\multicolumn{1}{c}{Ion} &
\multicolumn{1}{c}{Target}&
\multicolumn{1}{c}{Line}&
\multicolumn{1}{c}{}&
\multicolumn{1}{c}{}&
\multicolumn{1}{c}{Intensity}&
\multicolumn{1}{c}{}&
\multicolumn{1}{c}{}& 
\\
\cline{5-9}
\multicolumn{1}{c}{(eV)} & 
\multicolumn{1}{c}{} & 
\multicolumn{1}{c}{} &
\multicolumn{1}{c}{} &
\multicolumn{1}{c}{200 km/s} &
\multicolumn{1}{c}{400 km/s} &
\multicolumn{1}{c}{600 km/s} & 
\multicolumn{1}{c}{800 km/s} &
\multicolumn{1}{c}{1000 km/s}
}
\startdata
\hline
\\

298.96	&	C V	&	 N$_2$ & K$\alpha$ f 	&	0.6371	&	1.0257	&	1.2531	&	1.3689	&	1.4262 \\
304.41	&	C V	&	 N$_2$ & K$\alpha$ i 	&	0.0743	&	0.0901	&	0.0968	&	0.1007	&	0.1030 \\
307.90	&	C V	&	 N$_2$ & K$\alpha$ r 	&	1.0000	&	1.0000	&	1.0000	&	1.0000	&	1.0000 \\
354.52	&	C V	&	 N$_2$ & K$\beta$	&	3.1379	&	2.3855	&	1.9549	&	1.6592	&	1.4604 \\
370.92	&	C V	&	 N$_2$ & K$\gamma$	&	0.0011	&	0.0002	&	0.0001	&	0.0001	&	0.0000 \\
\\
\hline
\\ 
298.96	&	C V	&	 H$_2$O & K$\alpha$ f 	&	3.1488	&	2.9681	&	2.4518 	&	2.0284	&	1.7915 \\
304.41	&	C V	&	 H$_2$O & K$\alpha$ i 	&	0.2427	&	0.2299	&	0.1953 	&	0.1660	&	0.1479 \\
307.90	&	C V	&	 H$_2$O & K$\alpha$ r 	&	1.0000	&	1.0000	&	1.0000 	&	1.0000	&	1.0000 \\
354.52	&	C V	&	 H$_2$O & K$\beta$	&	0.4499	&	0.7120	&	1.2290	&	1.5967	&	1.7703 \\
370.92	&	C V	&	 H$_2$O & K$\gamma$	&	0.8395	&	0.7724	&	0.5980	&	0.4360	&	0.3298 \\
\\
\hline
\\
298.96	&	C V	&	 CO & K$\alpha$ f 	&	2.3460	&	1.0689	&	1.1178	&	1.2206	&	1.2946 \\
304.41	&	C V	&	 CO & K$\alpha$ i 	&	0.1977	&	0.1031	&	0.0977	&	0.0981	&	0.0992 \\
307.90	&	C V	&	 CO & K$\alpha$ r 	&	1.0000	&	1.0000	&	1.0000	&	1.0000	&	1.0000 \\
354.52	&	C V	&	 CO & K$\beta$		&	2.1029	&	2.6233	&	2.3414	&	2.0744	&	1.8629 \\
370.92	&	C V	&	 CO & K$\gamma$	&	0.3719	&	0.0653	&	0.0264	&	0.0153	&	0.0107 \\
\\
\hline
\\
298.96	&	C V	&	 CO$_2$ & K$\alpha$ f 	&	2.0166	&	1.1770	&	1.2407	&	1.3209	&	1.3720 \\
304.41	&	C V	&	 CO$_2$ & K$\alpha$ i 	&	0.1724	&	0.1065	&	0.1011	&	0.1010	&	0.1017 \\
307.90	&	C V	&	 CO$_2$ & K$\alpha$ r 	&	1.0000	&	1.0000	&	1.0000	&	1.0000	&	1.0000 \\
354.52	&	C V	&	 CO$_2$ & K$\beta$	&	2.1032	&	2.4034	&	2.1024	&	1.8275	&	1.6227 \\
370.92	&	C V	&	 CO$_2$ & K$\gamma$	&	0.3322	&	0.0718	&	0.0319	&	0.0194	&	0.0139 \\
\\
\hline
\\
298.96	&	C V	&	 O & K$\alpha$ f 	&	7.1031	&	11.7286	&	9.4888	&	6.8824	&	5.6041 \\
304.41	&	C V	&	 O & K$\alpha$ i 	&	0.5514	&	1.2741	&	1.1028	&	0.7808	&	0.6037 \\
307.90	&	C V	&	 O & K$\alpha$ r 	&	1.0000	&	1.0000	&	1.0000	&	1.0000	&	1.0000 \\
354.52	&	C V	&	 O & K$\beta$		&	0.3770	&	1.3046	&	2.4506	&	2.6876	&	2.6782 \\
370.92	&	C V	&	 O & K$\gamma$		&	1.1245	&	0.8812	&	0.4002	&	0.1916	&	0.1144 \\
\\
\hline
\\
298.96	&	C V	&	 OH & K$\alpha$ f 	&	5.7399	&	6.2376	&	4.8486	&	4.0528	&	3.7247 \\
304.41	&	C V	&	 OH & K$\alpha$ i 	&	0.3777	&	0.5407	&	0.4434	&	0.3534	&	0.3023 \\
307.90	&	C V	&	 OH & K$\alpha$ r 	&	1.0000	&	1.0000	&	1.0000	&	1.0000	&	1.0000 \\
354.52	&	C V	&	 OH & K$\beta$		&	0.3126	&	1.2453	&	2.0619	&	2.3243	&	2.3922 \\
370.92	&	C V	&	 OH & K$\gamma$	&	1.5362	&	1.1179	&	0.5853	&	0.3464	&	0.2395 \\
...	&	...	&	 ... & ...	&	...	&	...	&	...&	... &	... \\
...	&	...	&	 ... & ...	&	...	&	...	&	...&	... &	... \\

\enddata

\tabletypesize{\small}
\tablecomments{Line Ratios for X-ray emission resulting from C-Si bare and H-like ions CX collisions with N$_2$, H$_2$O, CO, CO$_2$, OH, and O.  Note that all lines are normalized to the Ly$\alpha$ line and K$\alpha$ resonsant line for bare ion and non-bare ion collisions, respectively.  Line ratios for bare ion collisions were obtained by applying the low energy distribution (Equation 3.2) to MCLZ $n$-resolved cross sections.  For non-bare ion collisions, MCLZ $n \ell S$-resolved cross sections were used to obtain these ratios.  See text for details.  Line ratios for bare ion collisions applying other distribution functions are available upon request.  See included machine readable table for complete C-Si CX data.}
\end{deluxetable*}

\newpage
\onecolumngrid
\LongTables
\begin{deluxetable}{c c c c c c}
\centering
\tabletypesize{\tiny}
\tablecaption{Comparison of Line Ratios for H-Like Emission}
\tablewidth{0pt}
\tablehead{
\multicolumn{1}{c}{Transition}&
\multicolumn{1}{c}{\cite{greenwood}}&
\multicolumn{1}{c}{MCLZ}&
\multicolumn{1}{c}{MCLZ}&
\multicolumn{1}{c}{MCLZ}&
\multicolumn{1}{c}{MCLZ}
\\
\multicolumn{1}{c}{}&
\multicolumn{1}{c}{}&
\multicolumn{1}{c}{Low Energy}&
\multicolumn{1}{c}{Modified Low Energy}&
\multicolumn{1}{c}{Separable}&
\multicolumn{1}{c}{Statistical}
}
\startdata
\hline
\\
\multicolumn{3}{l}{\underline{N$^{7+}$ + H$_2$O @ $\sim$3.3 keV/u}} \\
& & & & &
\\
Ly$\alpha$		& 1.000  & 1.000 & 1.000 & 1.000 & 1.000 \\
Ly$\beta$		& 0.762  & 0.363 & 0.203 & 0.128 & 0.125 \\
Ly$\gamma$ + Ly$\delta$	& 0.619  & 0.913 & 0.373 & 0.213 & 0.190 \\
Hardness Ratio 		& 1.381  & 1.276 & 0.576 & 0.341 & 0.315 \\
\\
\multicolumn{3}{l}{\underline{O$^{8+}$ + H$_2$O @ $\sim$3.1 keV/u}} \\
& & & & &
\\
Ly$\alpha$		& 1.000 & 1.000 & 1.000 & 1.000 & 1.000 \\
Ly$\beta$		& 0.130 & 0.324 & 0.184 & 0.112 & 0.106 \\
Ly$\gamma$ + Ly$\delta$	& 0.170 & 0.919 & 0.326 & 0.184 & 0.166 \\
Ly$\epsilon$ 		& 0.000 & 0.022 & 0.006 & 0.005 & 0.003 \\
Hardness Ratio 		& 0.300 & 1.265 & 0.516 & 0.301 & 0.275 \\
\\
\multicolumn{3}{l}{\underline{Ne$^{10+}$ + H$_2$O @ $\sim$3.2 keV/u}} \\
& & & & &
\\
Ly$\alpha$		& 1.000 & 1.000 & 1.000 & 1.000 & 1.000 \\
Ly$\beta$		& 0.123 & 0.283 & 0.170 & 0.091 & 0.084 \\
Ly$\gamma$ + Ly$\delta$	& 0.111 & 0.429 & 0.143 & 0.070 & 0.070 \\
Ly$\epsilon$ 		& 0.000 & 0.423 & 0.109 & 0.067 & 0.053 \\
Ly$\zeta$ 		& 0.000 & 0.045 & 0.010 & 0.008 & 0.005 \\
Hardness Ratio 		& 0.234 & 1.180 & 0.432 & 0.236 & 0.212 \\
\\
\multicolumn{3}{l}{\underline{O$^{8+}$ +CO$_2$ @ $\sim$3.1 keV/u}} \\
& & & & &
\\
Ly$\alpha$		& 1.000 & 1.000 & 1.000 & 1.000 & 1.000 \\
Ly$\beta$		& 0.132 & 0.362 & 0.207 & 0.123 & 0.122 \\
Ly$\gamma$ + Ly$\delta$	& 0.184 & 0.887 & 0.342 & 0.200 & 0.174 \\
Ly$\epsilon$ 		& 0.000 & 0.003 & 0.000 & 0.000 & 0.000 \\
Hardness Ratio 		& 0.316 & 1.252 & 0.549 & 0.323 & 0.296 \\
\\
\multicolumn{3}{l}{\underline{Ne$^{10+}$ +CO$_2$ @ $\sim$3.2 keV/u}} \\
& & & & &
\\
Ly$\alpha$		& 1.000 & 1.000 & 1.000 & 1.000 & 1.000 \\
Ly$\beta$		& 0.136 & 0.302 & 0.178 & 0.097 & 0.094 \\
Ly$\gamma$ + Ly$\delta$	& 0.010 & 0.565 & 0.199 & 0.098 & 0.100 \\
Ly$\epsilon$ 		& 0.000 & 0.314 & 0.082 & 0.056 & 0.040 \\
Ly$\zeta$ 		& 0.000 & 0.012 & 0.003 & 0.002 & 0.001  \\
Hardness Ratio 		& 0.146 & 1.193 & 0.462 & 0.253 & 0.235  \\
\\
\enddata

\tabletypesize{\normalsize}
\tablecomments{\\
The lines ratios presented for comparison in this table are taken from the \cite{greenwood} study.   Experimental line ratios from \cite{greenwood} are deduced from their reported relative contributions from the various transitions given above to the total spectrum.  Note that these data have been corrected for the transmission efficiency of the Be window utilized by the germanium X-ray detector.}
\end{deluxetable}\newpage
\onecolumngrid
\LongTables
\begin{deluxetable}{c c c}
\centering
\tabletypesize{\tiny}
\tablecaption{Comparison of Line Ratios for He-Like Emission}
\tablewidth{0pt}
\tablehead{
\multicolumn{1}{c}{Transition}&
\multicolumn{1}{c}{\cite{greenwood}}&
\multicolumn{1}{c}{MCLZ}
}
\startdata
\hline
\multicolumn{3}{l}{\underline{O$^{7+}$ + H$_2$O @ $\sim$2.7 keV/u}} \\
& &
\\
K$\alpha$	& 1.000  &	1.000 \\
K$\beta$	& 0.186  &	0.098 \\
K$\gamma$ + K$\delta$	& 0.243  &	0.437 \\
Hardness Ratio & 0.429 & 0.535 \\
\\
\multicolumn{3}{l}{\underline{Ne$^{9+}$ + H$_2$O @ $\sim$2.9 keV/u}} \\
& &
\\
K$\alpha$	& 1.000  &	1.000 \\
K$\beta$	& 0.033  &	0.112 \\
K$\gamma$	& 0.044 &	0.050 \\
K$\delta$	& 0.022 &	0.099 \\
K$\epsilon$ & 0.000 &  0.136\\
Hardness Ratio & 0.099 & 0.397 \\
\\
\multicolumn{3}{l}{\underline{O$^{7+}$ +CO$_2$ @ $\sim$2.7 keV/u}} \\
& &
\\
K$\alpha$	& 1.000  &	1.000 \\
K$\beta$	& 0.180 &	0.141 \\
K$\gamma$ + K$\delta$	& 0.192 &	0.507 \\
Hardness Ratio & 0.372 & 0.648 \\
\\
\multicolumn{3}{l}{\underline{Ne$^{9+}$ +CO$_2$ @ $\sim$2.9 keV/u}} \\
& &
\\
K$\alpha$	& 1.000  &	1.000 \\
K$\beta$	& 0.043  &	0.117 \\
K$\gamma$ & 0.022  &	0.069 \\
K$\delta$ & 0.022  &	0.176 \\
K$\epsilon$ & 0.000 & 0.074 \\
Hardness Ratio & 0.087 & 0.436 \\
\\
\enddata

\tabletypesize{\normalsize}
\tablecomments{\\
Same as Table 2 except comparing line ratios for He-like emission for collisions studied in \cite{greenwood}.   \cite{greenwood} was \textit{not} able to resolve the intercombination and resonance lines for He-like emission or observe the forbidden line, therefore, MCLZ line ratios are normalized to the sum of K$\alpha$ f, i, and r.}
\end{deluxetable}

\newpage

\onecolumngrid
\LongTables
\begin{deluxetable}{c c c c c}
\centering
\tabletypesize{\tiny}
\tablecaption{Comparison of $n$-resolved Cross Sections}
\tablewidth{0pt}
\tablehead{
\multicolumn{1}{c}{$\sigma_n$} &
\multicolumn{1}{c}{Exp.} (\%)&
\multicolumn{1}{c}{CTMC} (\%)&
\multicolumn{1}{c}{LZ$^a$ (\%)}&
\multicolumn{1}{c}{MCLZ (\%)}
}
\startdata
\hline
\\
\multicolumn{5}{l}{\underline{N$^{7+}$ + CO @ 2 keV/u}} \\
& & & &
\\
$\sigma_{n=6}$	 & -- &	2.00	 & -- &	--  \\
$\sigma_{n=5}$	 &76.20 $\pm$ 0.70 &	38.00 & 13.00	 &14.33 \\
$\sigma_{n=4}$	 &23.80 $\pm$ 0.90	&54.00 &78.00&78.45 \\
$\sigma_{n=3}$	 &-- &4.00	& 9.00 &	7.22 \\
& & & &
\\

\multicolumn{5}{l}{\underline{N$^{7+}$ + CO$_2$ @ 2 keV/u}} \\
& & & &
\\
$\sigma_{n=6}$ &	-- &	3.00 & -- & -- \\
$\sigma_{n=5}$	& 87.00 $\pm$ 	0.90 & 	40.00 &	14.00 & 15.29 \\
$\sigma_{n=4}$	& 13.00	$\pm$  0.80	& 52.00 & 	80.00	& 75.22 \\
$\sigma_{n=3}$	& -- & 4.00& 	6.00	& 9.49 \\
& & & &
\\

\multicolumn{5}{l}{\underline{N$^{7+}$ + H$_2$O @ 2 keV/u}} \\
& & & &
\\
$\sigma_{n=6}$	& 1.70 $\pm$ 0.60 &	4.00	& --	& -- \\
$\sigma_{n=5}$	& 86.20 $\pm$ 0.80	& 56.00 & 57.00 &	45.42 \\
$\sigma_{n=4}$ &	12.10 $\pm$ 0.60 &	36.00 &	43.00	& 53.28 \\
$\sigma_{n=3}$	& --& 3.00 &	-- &	1.30 \\
& & & &
\\

\multicolumn{5}{l}{\underline{O$^{7+}$ + CO @ 2 keV/u}} \\
& & & & 
\\
$\sigma_{n=6}$	& -- &-- &	-- &	-- \\
$\sigma_{n=5}$	& 74.90 $\pm$  0.50 & 	-- &	-- &	32.55 \\
$\sigma_{n=4}$	& 25.10 $\pm$ 0.70 &	-- &	-- &	66.68 \\
$\sigma_{n=3}$	 &	-- &	-- &	-- &	0.77 \\
& & & &
\\

\multicolumn{5}{l}{\underline{O$^{7+}$ + CO$_2$ @ 2 keV/u}} \\
& & & & 
\\
$\sigma_{n=6}$ &	-- &	-- &	-- &	-- \\
$\sigma_{n=5}$	 & 80.00 $\pm$  	0.70 & -- &	--	& 32.29 \\
$\sigma_{n=4}$ &	20.00 $\pm$ 	0.70 &	--	& -- &	66.14 \\
$\sigma_{n=3}$	 &	-- &	-- &	-- &	1.57 \\
& & & &
\\

\multicolumn{5}{l}{\underline{O$^{7+}$ + H$_2$O @ 2 keV/u}} \\
& & & & 
\\
$\sigma_{n=6}$ &	1.40 $\pm$ 	0.70 &	--	 & -- &	-- \\
$\sigma_{n=5}$	& 85.10 $\pm$  2.40 & -- &	-- &	79.25 \\
$\sigma_{n=4}$	& 13.50 $\pm$  1.20 & -- &	-- &	20.74 \\
$\sigma_{n=3}$ &	-- &	-- &	-- &	0.01 \\
& & & & 
\enddata

\tabletypesize{\normalsize}
\tablecomments{\\
$^a$: Landau-Zener (LZ) calculations similar to MCLZ.  See \cite{bigbook} for details. \\ \\
The data presented for comparison in this table are taken from the \cite{bigbook} study which gathers experimental data, CTMC calculations, and Landau-Zener calculations similar to those that are computed in this work. Note that cross sections are relative, and we report the $n$-resolved fraction as a percentage of the total cross section.}
\end{deluxetable}

\newpage
\onecolumngrid
\LongTables
\begin{deluxetable*}{c c c c c c c c c}
\centering
\tabletypesize{\tiny}
\tablecaption{Relative Difference between MCLZ H$_2$O CX Line Ratios and Bodweits et al. (2007)}
\tablewidth{0pt}
\tablehead{
\multicolumn{1}{c}{Phot. Energy} &
\multicolumn{1}{c}{Ion} &
\multicolumn{1}{c}{Line}&
\multicolumn{1}{c}{}&
\multicolumn{1}{c}{}&
\multicolumn{1}{c}{Rel. Diff.}&
\multicolumn{1}{c}{}&
\multicolumn{1}{c}{}& 
\\
\cline{4-9}
\multicolumn{1}{c}{(eV)} & 
\multicolumn{1}{c}{} &
\multicolumn{1}{c}{} &
\multicolumn{1}{c}{200 km/s} &
\multicolumn{1}{c}{400 km/s} &
\multicolumn{1}{c}{600 km/s} & 
\multicolumn{1}{c}{800 km/s} &
\multicolumn{1}{c}{1000 km/s}
}


\startdata

\hline
\\
298.96	&	C V		&	K$\alpha$ f	&	0.422	&	0.296	&	0.456	&	0.596	&	0.729 \\
304.41	&	C V		&	K$\alpha$ i	&	0.392	&	0.366	&	0.608	&	0.723	&	0.802 \\
307.90	&	C V		&	K$\alpha$ r	&	0.000	&	0.000	&	0.000	&	0.000	&	0.000 \\
354.52	&	C V		&	K$\beta$		&	0.381	&	1.001	&	1.445	&	1.539	&	1.505 \\
370.92	&	C V		&	K$\gamma$		&	0.733	&	1.113	&	1.055	&	0.956	&	0.820 \\
\\
\hline
\\
367.49	&	C VI	&	Ly$\alpha$		&	0.000	&	0.000	&	0.000	&	0.000	&	0.000 \\
435.55	&	C VI	&	Ly$\beta$		&	1.114	&	0.891	&	0.929	&	1.045	&	1.078 \\
459.37	&	C VI	&	Ly$\gamma$		&	1.307	&	1.178	&	1.122	&	1.207	&	1.239 \\
470.39	&	C VI	&	Ly$\delta$		&	1.575	&	1.868	&	1.927	&	1.910	&	1.879 \\
\\
\hline
\\
419.80	&	N VI	&	K$\alpha$ f	&	1.351	&	1.208	&	1.013	&	0.856	&	0.748 \\
426.32	&	N VI	&	K$\alpha$ i	&	0.627	&	0.630	&	0.559	&	0.492	&	0.463 \\
430.70	&	N VI	&	K$\alpha$ r	&	0.000	&	0.000	&	0.000	&	0.000	&	0.000 \\
497.97	&	N VI	&	K$\beta$		&	0.441	&	0.118	&	0.409	&	0.698	&	0.865 \\
521.58	&	N VI	&	K$\gamma$		&	1.042	&	1.209	&	1.267	&	1.344	&	1.386 \\
532.65	&	N VI	&	K$\delta$		&	1.756	&	1.895	&	1.913	&	1.881	&	1.986 \\
\\
\hline
\\
500.28	&	N VII	&	Ly$\alpha$		&	0.000	&	0.000	&	0.000	&	0.000	&	0.000 \\
592.93	&	N VII	&	Ly$\beta$		&	0.676	&	1.016	&	1.046	&	1.089	&	1.146 \\
625.35	&	N VII	&	Ly$\gamma$		&	1.195	&	1.514	&	1.478	&	1.424	&	1.404 \\
640.36	&	N VII	&	Ly$\delta$		&	0.823	&	1.246	&	1.263	&	1.309	&	1.343 \\
\\
\hline
\\
560.99	&	O VII	&	K$\alpha$ f	&	0.801	&	0.637	&	0.714	&	0.772	&	0.875 \\
568.59	&	O VII	&	K$\alpha$ i	&	0.100	&	0.005	&	0.103	&	0.168	&	0.285 \\
573.95	&	O VII	&	K$\alpha$ r	&	0.000	&	0.000	&	0.000	&	0.000	&	0.000 \\
665.62	&	O VII	&	K$\beta$		&	0.671	&	0.920	&	0.889	&	0.953	&	0.896 \\
697.80	&	O VII	&	K$\gamma$		&	0.880	&	1.415	&	1.513	&	1.523	&	1.487 \\
712.72	&	O VII	&	K$\delta$		&	1.037	&	1.486	&	1.556	&	1.608	&	1.608 \\
\\
\hline
\\
653.56	&	O VIII	&	Ly$\alpha$		&	0.000	&	0.000	&	0.000	&	0.000	&	0.000 \\
774.59	&	O VIII	&	Ly$\beta$		&	1.042	&	0.945	&	0.959	&	1.028	&	1.113 \\
816.95	&	O VIII	&	Ly$\gamma$		&	1.276	&	1.397	&	1.475	&	1.521	&	1.554 \\
836.55	&	O VIII	&	Ly$\delta$		&	1.529	&	1.486	&	1.477	&	1.531	&	1.565 \\
847.20	&	O VIII	&	Ly$\epsilon$	&	0.539	&	0.025	&	0.123	&	0.023	&	0.339 \\
\enddata

\tabletypesize{\small}
\tablecomments{Relative difference between MCLZ Line Ratios for CX collisions with H$_2$O and Bodewits et al.(2007) ratios for CX collisions with H (i.e. 2$\vert$ X$_{\textrm{Bodewits}}$ - X$_{\textrm{MCLZ}}$  $\vert$/ (X$_{\textrm{Bodewits}}$ + X$_{\textrm{MCLZ}}$)).  In many models, the latter ratios are applied to model cometary charge exchange no matter the target.  Differences between the ratios symbolize the large effect of target dependence in CX data.  All relative differences for the K$\alpha$ resonance (r) line and Ly$\alpha$ are zero because the hydrogen-like and helium-like emission data sets are normalized to these lines, respectively.  Theoretical calculations for bare ion collisions via MCLZ apply the low energy distribution.}
\end{deluxetable*}

\end{document}